\documentclass{PoS}

\title{1+1+1 flavor QCD+QED simulation\\ at the physical point}

\ShortTitle{1+1+1 flavor QCD+QED simulation at the physical point}

\author{
\speaker{N.~Ukita}${}^{    }$
\ \ for PACS-CS Collaboration
\\ \\
Center for Computational Sciences,
University of Tsukuba, Tsukuba, Ibaraki 305-8577, Japan
\\
E-mail: {\rm ukita@ccs.tsukuba.ac.jp}
}

\abstract{
We present our progress report on 1+1+1 flavor QCD+QED simulation at the  
physical point.
Calculations are carried out with 2+1 flavor QCD gauge configurations  
generated by the PACS-CS Collaboration. 
The dynamical QED effect and the up-down quark mass difference
are incorporated by the reweighting technique.
We also discuss some physics results.
}

\FullConference{XXIX International Symposium on Lattice Field Theory \\
		 July 10-16, 2011\\
		 Squaw Valley, Lake Tahoe, California}

\begin{document}

\section{Introduction}
It has been a long-standing goal to achieve lattice QCD simulations at the physical point. Recently in \cite{PACS-CS_1, PACS-CS_2}, we have reported the realization of this goal for the 2+1 flavor QCD case, where we have assumed isospin symmetry and have taken degenerate up and down quark masses, $m_u = m_d$. However, in nature isospin symmetry is broken by the light quark mass difference, $m_u \neq m_d$, and by the difference of the light quark electric charges, $Q_u \neq Q_d$. The isospin symmetry breaking causes mass splittings in isospin multiplets of light hadrons, {\it e.g.} $m_{K^0}-m_{K^\pm},\  m_{n}-m_{p}$.  The magnitude of splittings are not large and yet important since {\it e.g., } it is this difference which guarantees the stability of proton. Because measurements of physical observables of lattice simulations are improving beyond 1\% accuracy, we can, and should, take into account the isospin breaking effects.  Thus, our next target should be 1+1+1 QCD+QED lattice simulation at the physical point. 

A pioneering QCD+QED lattice simulation was carried out in \cite{Duncan_1, Duncan_2} using quenched approximation both for QCD and QED. In this work, the quark fields were explicitly coupled to the photon field which was superimposed on gluon fields generated in a conventional Monte Carlo simulation.  Several studies followed in quenched QED \cite{Namekawa}-\cite{Glaessle}, basically using the above method for including electromagnetic effects. The splittings of isospin multiplets of hadrons obtained in these quenched QED studies qualitatively matched the experimental ones.  Nonetheless, dynamical QED calculations were required for convincing results. The possibility of including dynamical QED effects were first investigated in \cite{Duncan_3} by computing the quark determinant ratio of QCD+QED to QCD on small lattices with the reweighting technique.  Since the stochastic method used for evaluating the reweighting factor required many random noises even on a small volume, and scaled as $V^2$ with lattice volume $V$, the calculation on realistically large volumes was expected to be difficult. For this reason, most previous attempts were made in QCD+quenched QED.

In this report, we present our preliminary results for 1+1+1 full QCD + full QED simulation at the physical point. The procedure is basically the one developed in \cite{Duncan_1, Duncan_3}. The 2+1 flavor QCD configurations on a $32^3\times 64$ lattice near the physical point generated by the PACS-CS Collaboration \cite{PACS-CS_2} are used as the base set for QCD+QED. The lattice action is the nonperturbatively ${\cal O}(a)$-improved Wilson quark action and the Iwasaki gauge action at $\beta=1.90$.  The lattice spacing is $a\sim0.1\, {\rm fm}$ and the volume is $(3\, {\rm fm})^3$ box. The reweighting technique is used for incorporating dynamical QED effects \cite{Duncan_3} and  the up-down quark mass difference, and for adjusting the quark masses to the physical point \cite{PACS-CS_2, Hasenfratz, RBC_UKQCD}.

\section{Test calculation of reweighting factor for electric charge}
In order to see the level of difficulty of calculating the reweighting factor for dynamical QED effects on QCD configurations with a stochastic method, we investigate the noise dependence of the reweighting factor for the strange quark electric charge from $eQ_s=0$ to $\frac{e_{\rm phys}Q_s}{100}$, where $e_{\rm phys}=\sqrt{\frac{4\pi}{137}}$ and $Q_s=-\frac{1}{3}$. QCD configurations were generated on a $32^3\times 64$ lattice \cite{PACS-CS_2}.
Photon fields are generated with a non-compact gauge field formulation treated with an appropriate gauge fixing and a treatment for zero modes as in \cite{Duncan_1, Blum_1, Hayakawa}. Employing random noises generated according to the Gaussian distribution, $\eta_i\ (i=1,\dots,N_\eta)$, the reweighting factor is evaluated as,
\begin{eqnarray}
 {\rm Det}\, W =
 \lim_{N_\eta\rightarrow \infty}
 \frac{1}{N_\eta} \sum_{i=1}^{N_\eta} 
  {\rm e}^{-\eta^{\dagger}_i(W^{-1}-1)\eta_i},
 \ \ \ \ \ \
 W = \frac{D(\frac{e_{\rm phys}Q_s}{100}, \kappa_s)}{D(0, \kappa_s)},
\label{eq1}
\end{eqnarray}
where $D(eQ_S, \kappa_s)$ is the Dirac matrix with an electric charge $eQ_s$ and a hopping parameter $\kappa_s$, $W$ is the ratio of the Dirac matrices. 

\begin{figure}[t!]
\begin{center}
\begin{tabular}{cc}
\includegraphics[width=70mm,angle=0]{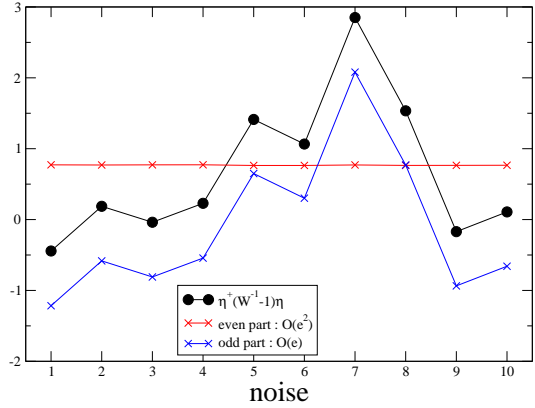}
\end{tabular}
\end{center}
\vspace{-.5cm}
\caption{Noise dependence of exponent of reweighting factors for the strange quark electric charge from $eQ_s=0$ to $(e_{\rm phys}Q_s)/100$, where $e_{\rm phys}=\sqrt{4\pi/137}$, $Q_s=-1/3$. Black filled circles show the exponents, $\eta^{\dagger}_i(W^{-1}-1)\eta_i$, and have large fluctuation. The even part of the exponents for the electric charge (red crosses) is stable. Fluctuation comes mainly from the odd part (blue crosses). }
\label{fig1}
\end{figure}

Figure \ref{fig1} shows the exponent of the reweighting factor $\eta^{\dagger}_i(W^{-1}-1)\eta_i$ for 10 noises labeled by $i=1,2,\cdots, 10$.  We see a strong noise dependence (see black filled circles) even though the electric charge is turned on at only 1\% of the physical value. 
It is important to observe that the main fluctuations come from the ${\cal O}(e)$ odd part in the electric charge (blue crosses in Fig. \ref{fig1}).  The ${\cal O}(e^2)$ even part (red crosses in Fig. \ref{fig1}) is quite stable. The odd part is not generally forbidden by symmetries of the QCD+QED mixed system, so cannot be removed by hand.

To reduce (or cancel) the ${\cal O}(e)$ odd part, we estimate the reweighting factor for all three light quarks simultaneously for each noise \cite{Izubuchi}, 
\begin{eqnarray}
 {\rm Det}\, W =
 \lim_{N_\eta\rightarrow \infty}
 \frac{1}{N_\eta} \sum_{i=1}^{N_\eta} 
  {\rm e}^{-\eta^{\dagger}_i(W^{-1}-1)\eta_i},
 \ \ \ \ \ \
 W = 
 \frac{D(q_u^{\prime}, \kappa_u)}{D(q_u, \kappa_u)}
 \frac{D(q_d^{\prime}, \kappa_d)}{D(q_d, \kappa_d)}
 \frac{D(q_s^{\prime}, \kappa_s)}{D(q_s, \kappa_s)},
\label{eq2}
\end{eqnarray}
where $q_{f} = e Q_{f}$ and $q_{f}^{\prime} = e^{\prime} Q_{f}\ (f=u,d,s)$.
Since the total charge for up, down and strange quarks vanishes, $q_u+q_d+q_s=0$, we can expect the cancelation of the odd part, ${\cal O}(e)=0$, if the quark masses are identical.  Even though the degeneracy is broken in nature, we expect that ${\cal O}(e)$ terms are reduced and so are the fluctuations of the reweighting factor from them. 
 
To reduce the noise dependence further, we modify the generation of photon fields as follows.
\begin{itemize}
\item[1)]
Photon fields are first generated on a $64^3\times128$ lattice which is twice finer than the QCD lattice of $32^3\times64$, by using a non-compact gauge action with an appropriate gauge fixing and with a treatment for zero modes as in \cite{Duncan_1, Blum_1, Hayakawa}. 
\item[2)]
To reduce short distance fluctuations of photon fields on QCD lattice, we average the photon fields over all independent paths inside the $2^4$ cell of the QED lattice, which corresponds to the unit-cell of QCD lattice. 
\item[3)]Finally, compact link fields $U_{\mu}$ coupling to quarks with electric charges $e_{\rm phys}Q$ are defined by exponentiating the averaged photon fields $A_{\mu}$; $U_{\mu} = {\rm e}^{ie_{\rm phys}QA_{\mu}}$. 
\end{itemize}
In this procedure, we neglect the running of the electric charge, {\it i.e.,} take always $e_{\rm phys} = \sqrt{\frac{4\pi}{137}}$.
We numerically checked that the averaging procedure did not cause running of the electric charge within error bars, by considering electromagnetic mass splittings between charged and neutral pseudoscalar mesons in the flavor non-singlet sector and comparing with the case of conventional photon fields. 
We found that the use of the averaged fields reduced the noise   dependence of the reweighting factor.  This procedure also reduced a large additive quark mass shift for Wilson quark action due to QED effects to a small one.  We use these link fields as photons from now on.

\section{Method for 1+1+1 flavor QCD+QED simulation at the physical point}
We carry out 1+1+1 flavor QCD+QED simulation at the physical point as follows,
\begin{itemize}
 \item[1)] use 2+1 flavor QCD configurations near the physical point with the nonperturbatively ${\cal O}(a)$-improved Wilson quark action and the Iwasaki gauge action at $\beta=1.90$ on a $32^3\times64$ lattice generated by the PACS-CS Collaboration,
 \item[2)] generate photon fields as described in the previous section, then superimpose the photon fields on the QCD configurations,
 \item[3)] reweight into 1+1+1 flavor QCD+QED action with different hoping parameters $\kappa_{u,d,s}$ and the physical electric charge $e_{\rm phys}$ to incorporate dynamical QED effects,
 \item[4)] tune three hopping parameters to the physical point values defined by $\pi^+, K^+, K^0$ and $\Omega^-$ masses as the physical input.
\end{itemize}
For the calculation of the reweighting factor, we introduce a square root trick to cut down on the calculation time, ${\rm Det}\, W = \sqrt{({\rm Det}\, W)^2}$, and determinant breakups to reduce the noise fluctuation \cite{PACS-CS_2, Hasenfratz, RBC_UKQCD}. The number of determinant breakups used is 438. 12 noises are used for each determinant breakup. We also use the technique of calculating the effects of the up, down and strange electric charges simultaneously for each noise as explained in Eq. (\ref{eq2}). The inversion algorithm for Dirac matrices is the block solver of \cite{Nakamura} which accelerates by a factor of $3\sim4$ compared with non-block solvers. 

\section{Preliminary results}
In this section, we present our preliminary results on 1+1+1 flavor QCD+QED simulation at the physical point. So far, the number of configurations analyzed is 35 out of 80. 

\begin{figure}[t!]
\begin{center}
\begin{tabular}{cc}
\includegraphics[width=75mm,angle=0]{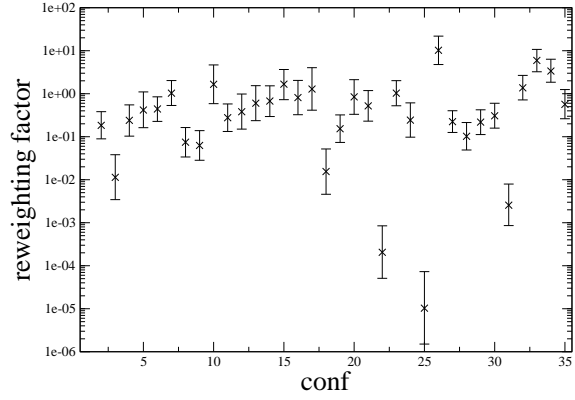}
\end{tabular}
\end{center}
\vspace{-.5cm}
\caption{Configuration dependence of the reweighting factor from 2+1 flavor QCD to 1+1+1 flavor QCD+QED.}
\label{fig2}
\end{figure}

Fig. \ref{fig2} shows the configuration dependence  of the reweighting factor from 2+1 flavor QCD to 1+1+1 flavor QCD+QED at the physical point. Except for a few configurations, the fluctuation is contained within ${\cal O}(10)$ due to the introduction of the techniques as described in the previous sections.

\begin{figure}[t!]
\begin{center}
\begin{tabular}{c}
\includegraphics[width=60mm,angle=0]{fig3_1.eps}
\end{tabular}
\end{center}
\end{figure}
\begin{figure}[t!]
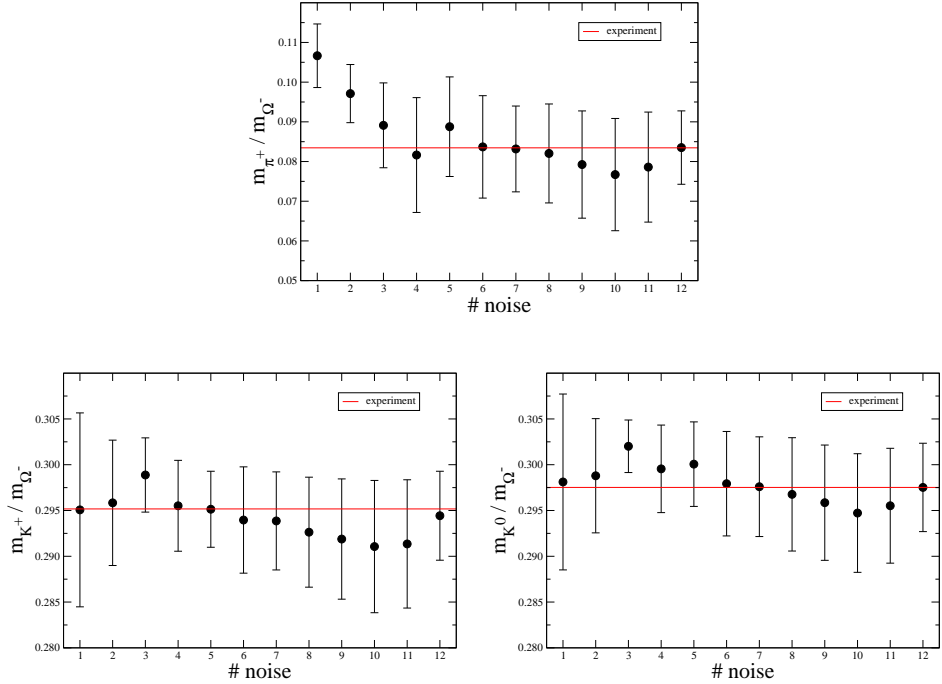

\vspace{-3mm}
\begin{center}
\begin{tabular}{cc}
\includegraphics[width=60mm,angle=0]{fig3_2.eps}
&
\includegraphics[width=60mm,angle=0]{fig3_3.eps}
\end{tabular}
\end{center}
\vspace{-.5cm}
\caption{Noise dependence of $m_{\pi^+}, m_{K^+}$ and $m_{K^0}$ normalized by $m_{\Omega^-}$ (black filled circles). Red lines indicate the experimental values. The horizontal lines indicate the number of noises used in each determinant breakup for the reweighting factor.}
\label{fig3}
\end{figure}

For hadron measurements, four source points were used. We tuned the hopping parameters to the physical point values using  $\pi^+, K^+, K^0$ and $\Omega^-$ masses as the physical input. 
Figure \ref{fig3} shows the hadron mass ratios, $\frac{m_{\pi^+}}{m_{\Omega^-}}, \frac{m_{K^+}}{m_{\Omega^-}}$ and $\frac{m_{K^0}}{m_{\Omega^-}}$ (black filled circles), as a function of the number of noises used for each determinant breakup. The ratios reach a plateau around  the experimental values (red lines) already from small number of noises.  This shows that twelve noises for each determinant breakup is sufficient for the reweighting factor calculation.  Figure \ref{fig3} also shows that the tuning to the physical point has been achieved up to error bars.

\begin{figure}[t!]
\begin{center}
\begin{tabular}{cc}
\includegraphics[width=75mm,angle=0]{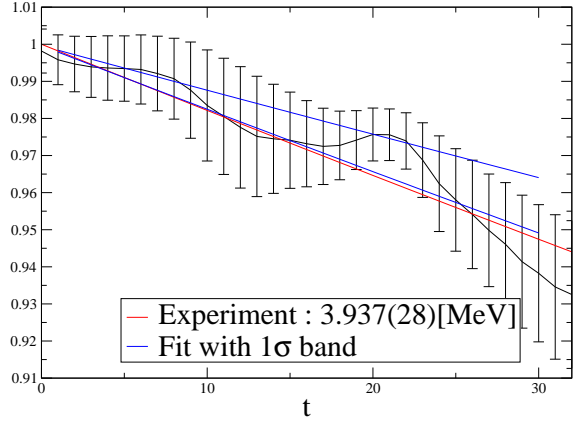}
\end{tabular}
\end{center}
\vspace{-.5cm}
\caption{Ratio of $K^0$ to $K^+$ propagators which shows the mass difference $m_{K^0}-m_{K^+}$. Black curve is the lattice result which is roughly consistent with the experimental values (red curve). Blue curves are fit ones with 1$\sigma$ error band.}
\label{fig4}
\end{figure}

Figure \ref{fig4} shows the ratio of $K^0$ to $K^+$ propagators which is  expected to behave as a function of time $t$ as,
\begin{eqnarray}
 \frac{\left\langle K^0(t)K^0(0) \right\rangle}{\left\langle K^+(t)K^+(0)\right\rangle}
   &\propto&
    1 - (m_{K^0}-m_{K^+})t, 
\end{eqnarray}
where we assume $(m_{K^0}-m_{K^+})t \ll 1$. The slope indicates the mass difference. The lattice data (black curve) and the fit result with $1\sigma$ error band 3.21(57)\, MeV (blue curve) are roughly consistent with the experimental value 3.937(28)\, MeV \cite{Exp} (red curve). 

Finally in Table 1, we summarize the up, down and strange quark masses renormalized at $\mu = 2$\, GeV in the continuum $\overline{\rm MS}$ scheme. The renormalization factor is nonperturbatively determined in the Schr\"odinger functional scheme \cite{PACS-CS_3}.  Here, we neglect QED corrections to the renormalization factor.

The results presented here were obtained from 35 configurations out of 80 ones and are still preliminary. In order to conclude 1+1+1 flavor QCD+QED simulation at the physical point, we are now increasing the statistics and also source points for measurements to reduce the statistical errors.

\begin{table}[h!]
\begin{center}
\renewcommand{\arraystretch}{1.1}
\begin{tabular}{lcl}
\hline\hline
$m_{u}$ &=& 1.97(67) [MeV]\\
$m_{d}$ &=& 4.31(83) [MeV]\\
$m_{s}$ &=& 90.32(67) [MeV]\\
$(m_{u}+m_d)/2$ &=& 3.14(72) [MeV]\\
$m_{u}/m_{d}$ &=& 0.457(93)\\
$2 m_{s}/(m_u + m_d)$ &=& 28.8(6.6)\\
\hline\hline
\end{tabular}
\end{center}
\caption{Preliminary results for renormalized quark masses in 1+1+1 flavor QCD+QED at $\beta=1.90$ on $32^3\times64$ lattice. $\mu = 2\, {\rm GeV}$.}
\label{default}
\end{table}%

\section*{Acknowledgments}
This work is supported in part by Grants-in-Aid of the Ministry
of Education, Culture, Sports, Science and Technology-Japan
 (Nos. 18104005, 20340047,
 20540248, 21340049, 22244018, and 22740138),
and Grant-in-Aid for Scientific Research on Innovative Areas
(No. 2004: 20105001, 20105002, 20105003, 20105005, and
 No. 2104: 22105501).
The numerical calculations have been carried out
on T2K-Tsukuba and PACS-CS at Center for Computational Sciences, University of Tsukuba.



\begin{thebibliography}{99}
 \bibitem{PACS-CS_1} S. Aoki {\it et al.} [PACS-CS Collaboration], Phys. Rev. {\bf D79}, 034503 (2009). 
 
 \bibitem{PACS-CS_2} S. Aoki {\it et al.} [PACS-CS Collaboration], Phys. Rev. {\bf D81}, 074503 (2010).  
 
 \bibitem{Duncan_1} A. Duncan, E. Eichten and H. Thacker, Phys. Rev. Lett. {\bf 76}, 3894 (1996).
 
 \bibitem{Duncan_2} A. Duncan, E. Eichten and H. Thacker, Phys. Lett. {\bf B409}, 387 (1997).
 
 \bibitem{Namekawa} Y.~Namekawa and Y.~Kikukawa, PoS {\bf LATTICE2005}, 090 (2006).

 \bibitem{Yamada}
  N.~Yamada, T.~Blum, M.~Hayakawa and T.~Izubuchi  [RBC Collaboration],
  PoS {\bf LATTICE2005}, 092 (2006).

 \bibitem{Doi}
  T.~Doi, T.~Blum, M.~Hayakawa, T.~Izubuchi and N.~Yamada,
  PoS {\bf LATTICE2006}, 174 (2006).

 \bibitem{Blum_1} T. Blum, T. Doi, M. Hayakawa, T. Izubuchi and N. Yamada, Phys. Rev. {\bf D76}, 114508 (2007). 
 
 \bibitem{MILC_1} S. Basak {\it et al.} [MILC Collaboration], PoS {\bf LATTICE2008}, 127 (2008). 

 \bibitem{Zhou}
  R.~Zhou, T.~Blum, T.~Doi, M.~Hayakawa, T.~Izubuchi and N.~Yamada,
  PoS {\bf LATTICE2008}, 131 (2008).
 
 \bibitem{Zhou:2009ku}
  R.~Zhou and S.~Uno,
  PoS {\bf LATTICE2009}, 182 (2009).

 \bibitem{Blum_2} T. Blum, R. Zhou, T. Doi, M. Hayakawa, T. Izubuchi, S. Uno and N. Yamada, Phys. Rev. {\bf D82}, 094508 (2010). 
 
  \bibitem{BMW_1} A. Portelli {\it et al.} [Budapest-Marseille-Wuppertal Collaboration], PoS {\bf LATTICE2010}, 121 (2010).

 \bibitem{MILC_2} A. Torok {\it et al.} [MILC Collaboration], PoS {\bf LATTICE2010}, 127 (2010). 
 
 \bibitem{BMW_2} A. Portelli, PoS {\bf LATTICE2011}, 136 (2011).

 \bibitem{Glaessle} B. Gl\"a\ss le, PoS {\bf LATTICE2011}, 282 (2011).
    
 \bibitem{Duncan_3} A. Duncan, E. Eichten and R. Sedgewick, Phys. Rev. {\bf D71}, 0945509 (2005).
 
 \bibitem{Hasenfratz} A. Hasenfratz, R. Hoffmann and S Shaefer, Phys. Rev. {\bf D78}, 014515 (2008).
 
 \bibitem{RBC_UKQCD} Y. Aoki {\it et al.} [RBC Collaboration and UKQCD Collaboration], Phys. Rev. {\bf D83}, 074508 (2011).
 
 \bibitem{Hayakawa}
  M.~Hayakawa and S.~Uno,
  Prog.\ Theor.\ Phys.\  {\bf 120}, 413 (2008).
   
 \bibitem{Izubuchi} Private communication with T. Izubuchi.

 \bibitem{Nakamura} Y.~Nakamura, K.~I.~Ishikawa, Y.~Kuramashi, T.~Sakurai and H.~Tadano, Comput. Phys. Commun. {\bf 183}, 34 (2012).
 
 \bibitem{Exp} K. Nakamura {\it et al.} [Particle Data Group], J. Phys. G {\bf 37}, 075021 (2010).

 \bibitem{PACS-CS_3} S. Aoki {\it et al.} [PACS-CS Collaboration], JHEP {\bf 1008}, 101 (2010).
 
\end{thebibliography}
\end{document}